\documentclass[twocolumn,superscriptaddress,amsmath,amssymb,aps,prb]{revtex4-1}

\usepackage{graphicx}
\usepackage[space]{grffile}
\usepackage{dcolumn}
\usepackage{bm}
\usepackage{subfigure}
\usepackage{mathrsfs}
\usepackage{booktabs}
\usepackage{notoccite}
\usepackage{float}
\usepackage{placeins}
\usepackage{siunitx}
\usepackage{physics}

\DeclareSIUnit{\angstrom}{\mbox{\normalfont\AA}}
\newcommand*{\medcap}{\mathbin{\scalebox{1.8}{\ensuremath{\cap}}}}%

\begin{document}


\title{DiSPy: Implementation of Distortion Symmetry for the Nudged Elastic Band Method}

\author{Jason M. Munro}
\email{munrojm@psu.edu}
\affiliation{Department of Materials Science and Engineering and Materials Research Institute, The Pennsylvania State University, University Park, PA 16802, USA}
\affiliation{Penn State Institutes of Energy and the Environment, The Pennsylvania State University, University Park, PA 16802, USA}

 \author{Vincent S. Liu}
 \affiliation{Department of Materials Science and Engineering and Materials Research Institute, The Pennsylvania State University, University Park, PA 16802, USA}
 
\author{Venkatraman Gopalan}
\affiliation{Department of Materials Science and Engineering and Materials Research Institute, The Pennsylvania State University, University Park, PA 16802, USA}
\affiliation{Department of Physics, The Pennsylvania State University, University Park, Pennsylvania 16802, USA}
\affiliation{Department of Engineering Science and Mechanics, The Pennsylvania State University, University Park, Pennsylvania 16802, USA}

\author{Ismaila Dabo}
\affiliation{Department of Materials Science and Engineering and Materials Research Institute, The Pennsylvania State University, University Park, PA 16802, USA}
\affiliation{Penn State Institutes of Energy and the Environment, The Pennsylvania State University, University Park, PA 16802, USA}


\begin{abstract}
The nudged elastic band (NEB) method is a commonly used approach for the calculation of minimum energy pathways of kinetic processes. However, the final paths obtained rely heavily on the nature of the initially chosen path. This often necessitates running multiple calculations with differing starting points in order to obtain accurate results. Recently, it has been shown that the NEB algorithm can only conserve or raise the distortion symmetry exhibited by an initial pathway. Using this knowledge, symmetry-adapted perturbations can be generated and used as a tool to systematically lower the initial path symmetry, enabling the exploration of other low-energy pathways that may exist. The group and representation theory details behind this process are presented and implemented in a standalone piece of software (\texttt{DiSPy}). The method is then demonstrated by applying it to the calculation of ferroelectric switching pathways in LiNbO$_3$. Previously reported pathways are more easily obtained, with new paths also being found which involve a higher degree of atomic coordination.
\end{abstract}

\maketitle


\section{Introduction}
A distortion is the pathway by which a system transforms in its transition between two or more physical states. Distortions are ubiquitous in nature and crucial in our understanding of physical processes. Recently, the concept of distortion symmetry, an analog of the symmetry of static structures, has been introduced by VanLeeuwen and Gopalan\cite{VanLeeuwen2015} as a framework to study distortions. The versatility of distortion symmetry is shown through an array of insightful examples and valuable applications, including molecular distortions and pseudorotations, minimum energy paths, tensor properties of crystals, and electronic structures and Berry phases \cite{VanLeeuwen2015}. The application to the discovery of minimum energy paths (MEPs) was further explored by Munro \textit{et. al.}\cite{Munro2018}, where MEPs overlooked by conventional techniques are consistently and systematically generated through symmetry-adapted perturbations made possible by distortion symmetry. These results agree with and extend MEPs found previously by other researchers using approaches involving physical intuition and single structure perturbations.

We introduce a thorough treatment of distortion symmetry and its application to nudged elastic band (NEB) calculations employed by Munro \textit{et al.}\cite{Munro2018}, providing a complete introduction to the concepts, mathematics, and methods of distortion symmetry. Furthermore, we illustrate its synergy with powerful methods in representation theory for generating symmetry-adapted perturbations in the context of materials science. In particular, we show the application of our method to ferroelectric switching in LiNbO$_3$ using a software implementation (\texttt{DiSPy})\cite{dispy}. 

\section{Theory and Background}
\subsection{Distortion Symmetry Groups}
Consider a periodic cell of a crystalline solid in three-dimensions which contains $N$ atoms. This cell can be represented by the collection of vectors $V = \{\mathbf{r}_i^\alpha\ |\ i=1,\ldots,N\}$, where $\mathbf{r}_i^\alpha$  give the fractional coordinate of atom $i$ with type $\alpha$ in the cell basis. The spatial symmetry of the structure described by this cell is characterized by the group ($S$) of conventional spatial operations ($h \in S$) represented by matrix-vector pairs $(R,\mathbf{t})$, which leave $V$ invariant:
\begin{equation}
	R\mathbf{r}_{i}^{\alpha}+\mathbf{t}=\mathbf{r}_{i'}^\alpha\in V,
\end{equation}
for all $h$ and $\mathbf{r}_i^\alpha$. Here, $R$ is the matrix representation of the proper or improper rotation associated with $h$, and $\mathbf{t}$ is the spatial translation vector. 

\begin{figure}
	\centerline{\includegraphics[width=\linewidth]{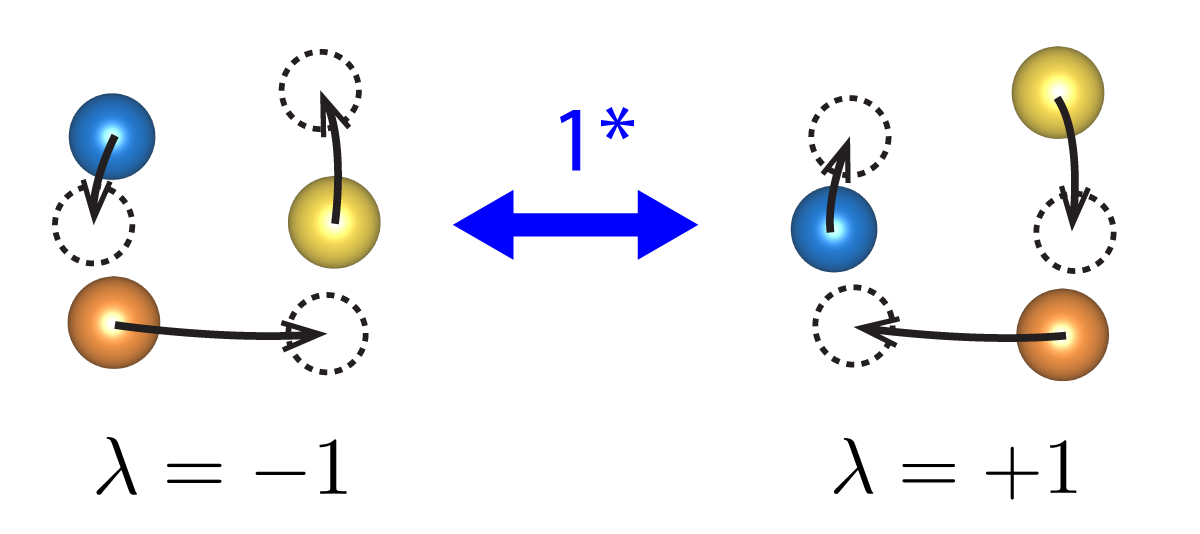}} 
	\caption{A set of displacements of a collection of atoms constituting a distortion. The distortion is paramaterized by a reaction coordinate $\lambda$, which varies between $\lambda=-1$ and $\lambda=+1$. Applying the $1^*$ antisymmetry operation reverses the distortion by taking $\lambda \rightarrow -\lambda$.\label{dsym}}
\end{figure}

Whereas conventional symmetry groups represent the symmetry of a static structure, distortion symmetry and distortion symmetry groups characterize the symmetry of a distortion pathway. By considering a pathway described by a set of structures $P = \{V_m\}$, the distortion symmetry group ($G$) of a given distortion pathway is the closed set of operations $(G=\{g\})$ which, when applied to said pathway, leave it unchanged or result in an equivalent pathway configuration:
\begin{equation}
R\mathbf{r}_{i,m}^{\alpha}+\mathbf{t}=\mathbf{r}_{i',m'}^\alpha\in P,
\end{equation}
for all $g$ and $\mathbf{r}_{i,m}^\alpha$. When considering the transformations appropriate for distortion symmetry, VanLeeuwen and Gopalan \cite{VanLeeuwen2015} first note that a distortion is described by the positions of all atoms in space at every point in the distortion. It is thus parameterized by spatial coordinates and a time-like dimension representing the progress or extent of the distortion, both of which should be operated upon. To operate on space (independent of progress/extent), the elements of conventional symmetry groups are applied to entire distortion pathways by transforming all atoms undergoing the distortion at every point in the path. To operate on the extent of distortion, a new operation known as distortion reversal ($1^*$) is introduced, which reverses the entire distortion and generates a pathway from the final state to the initial state (Figure \ref{dsym}). $1^*$ can be combined with conventional symmetry operations, and as such, any element in a distortion symmetry group is an element of the direct product of the group of all conventional symmetry elements and the group $\{1,1^*\}$.
However, merely referring to the ``extent" of a distortion is problematic. One can define the extent of any point in a physical distortion in any number of ways (for instance, the elapsed time or the position of one atom), each method potentially yielding different results when distortion reversal is applied. VanLeeuwen and Gopalan\cite{VanLeeuwen2015} introduce a parameter ($\lambda$) as a reaction (or distortion) coordinate which is consistent and universal. 

As $1^*$ commutes with and is different from any element of a conventional symmetry group, every element of a distortion symmetry group is either mathematically equivalent to a conventional symmetry operation or can be uniquely decomposed into the product of $1^*$ and a conventional symmetry operation. Those elements that fall into the former group are denoted ``unstarred" and those that fall into the latter group ``starred."

In nature, distortions are continuous -- they contain an uncountably infinite number of intermediate states. Given a continuous definition of the distortion parameter $\lambda$, it is possible to determine the distortion symmetry group of a path. However, this is difficult and unnecessary, as a smooth path can be approximated very precisely with discrete snapshots, or images, of the path. We will focus on determining distortion symmetry given only the set of these images $\{V_\lambda\}$. As they do not contain the entirety of the information of the path, we must reformulate the terms of distortion symmetry to best suit this approximation. 

It would be most precise and convenient if the images are evenly spaced in $\lambda$. This turns out to be the case for images in NEB calculation, due to the nature of the simulated spring forces at work (assuming images are near enough to each other to be approximated by differentials). With this, applying distortion reversal to a path results in the new distortion coordinate of every image coinciding with the initial distortion coordinate of its ``opposite'' image at $-\lambda$. Thus, distortion reversal implies, quite simply, mapping the first image to the distortion coordinate of last, the second to the second to last, and so forth. Furthermore, it is also preferable to work with an odd number of images; if a middle image exists, it is often a critical transition image. Moreover, distortion reversal also leaves the distortion coordinate of the middle image unchanged, providing a bridge between the conventional symmetry of the middle image and the distortion symmetry of the path. 

With this framework, it is now rather simple to construct the group of distortion symmetry operations from a set of appropriate images $\{V_\lambda\}$ parameterized with $\lambda$ corresponding to a distortion pathway ($P$).\cite{VanLeeuwen2015} First, the operations are separated into the unstarred and starred categories. If an element is unstarred, it must leave all images in the distortion pathway invariant. Thus, these elements are able to be obtained through a simple intersection of the conventional symmetry groups of all images. Let us denote all such elements as $H$, which can be written as
\begin{equation}
H = \medcap_{_{-1\leq\lambda\leq+1}} S(\lambda),
\end{equation}
where $S(\lambda)$ is the conventional symmetry group of the image $V_\lambda$. $H$ is a group, as it is the intersection of groups. If an element is starred, its spatial component must be a symmetry operation of the middle image $V_0$. These elements can be determined by finding all symmetry operations of the middle image that map images at $\lambda$ to those at $-\lambda$ $(V_\lambda \rightarrow V_{-\lambda})$. The set of all such operations is denoted as $A$.

Seeing that all of the elements of a distortion group belong to one of these two categories, we can write the distortion group ($G$) as:
\begin{equation}
G = H \cup 1^*A
\end{equation}

It is also very convenient that all distortion symmetry groups are isomorphic to at least one space group, and thus have identical irreducible representations (irreps) \cite{Vanleeuwen2014}. We first note that $1^*$ is not an element of any physically significant distortion symmetry group: if the opposite were true, the path could be broken down into two identical distortions, which could be analyzed in place of the original. Any distortion group $G=H \cup A^*$ is isomorphic to the set $G'=H\cup A^*1^*$, as $1^*$ is not an element of $A^*$. Furthermore, it is simple to show that $G’$ is closed and thus is a valid space group. One can thus generate a space group isomorphic to any physically significant distortion symmetry group by making all of its starred elements unstarred.
\subsection{Perturbing Initial Paths}
We now turn to the application of distortion symmetry to the nudged elastic band method (NEB) employed in the paper by Munro \textit{et al.}\cite{Munro2018}. NEB calculations have the potential to neglect a number of potential minimum energy pathways. This can be thought of as an analogue to the common failure of static structure relaxation calculations in providing the ground-state due to force balancing and symmetry conservation. However, instead of static symmetry being conserved, it is instead the distortion symmetry of a path\cite{VanLeeuwen2015}. In the case of static structure calculations, the issue can be solved by perturbing the generated structure with unstable phonon modes to lower its symmetry. This, however, is computationally costly and becomes exponentially more so in the case of entire paths, making a calculation of unstable modes of a path infeasible in many circumstances. Instead of this, we can utilize the related and less computationally intensive subject of distortion symmetry. Using distortion groups, basis vectors for unstable modes of a path can be generated and used as alternative perturbations. Since the NEB algorithm is unable to lower the distortion symmetry of the initial pathway\cite{VanLeeuwen2015}, generating perturbations that selectively break distortion symmetry and invoke path instabilities useful in enabling NEB to explore additional pathways that may exist. 

\subsubsection{Distortion space and irreducible representations}

Consider the vector space $U$, hereon referred to as ``distortion space", of all possible distortions whose initial and final images are identical to those of the initial path, with addition and scalar multiplication defined as the addition and scalar multiplication of the vector displacements of every atom at every $\lambda$ between $-1$ and $1$. A vector in this space ($\mathbf{v}\in U$) can be written as a linear combination of $3Np$ basis vectors of the space, where $p$ is the number of images in the path, and $N$ is the number of atoms in each image. In other words,
\begin{equation}
\mathbf{v} = \sum_{i=1}^{3N}\sum_{j=1}^{p}C_i^j\mathbf{d}_{i}^j
\end{equation}
where $C_i^j$ is a constant, $\mathbf{d}_{i}^j$ is a basis vector of $U$ representing a displacement of one of the $N$ atoms in image $j$ in one of the three cell vector direction, and for all $i$,
\begin{equation}
|\mathbf{d}_{i}^1|, |\mathbf{d}_{i}^p| = 0.
\end{equation}
It is simple to show that applying a perturbation (adding an vector in distortion space to the $3Np$-dimensional vector constructed with $\mathbf{r}_{i,m}^\alpha$ vectors that represents the original path) preserves all distortion symmetry operations shared between the initial path and the perturbation, and breaks all symmetry operations of the initial path that are not symmetry operations of the perturbation. To control the distortion symmetry of the perturbed path, we should therefore control the distortion symmetry of the perturbation.

Suitable perturbations can be obtained by considering the irreducible representations (irreps) of the distortion symmetry group of the initial path. For a brief description of irreps, see Supplementary Note 1. Similar to the way phonon modes in a crystal can be categorized using the irreps of its space group, distortion group irreps allow us to categorize displacive modes of atoms in a path. Since $U$ is invariant with respect to the elements $g \in G$,
\begin{equation}
g \mathbf{d}_{i}^j = \sum_{i=1}^{3N}\sum_{j=1}^{p}D^{U}_{iji'j'}(g)\mathbf{d}_{i}^j =  \mathbf{d}_{i'}^{j'} \in U,
\end{equation}
for all $g$ and $\mathbf{d}_{i}^j$. Where $D^{U}(g)$ is the matrix of $g$ in the reducible representation $D^{U}$ constructed using $U$, it is then possible to obtain a set of distortion vectors adapted to the irreps of $G$. Mathematically, this corresponds to generating basis vectors of symmetry invariant subspaces of the distortion space ($W_\mu \subset U$):
\begin{equation}
g\mathbf{w}_{i}^{\mu} = \sum_{i=1}^{l_n}D^{\mu}_{ii'}(g)\mathbf{w}_{i}^{\mu} =  \mathbf{w}_{i'}^{\mu} \in W_\mu,
\end{equation}
for all $g$, and $\mathbf{w}_{i}^{\mu}$. Where $D^{\mu}(g)$ is the matrix of $g$ in the irreducible representation $D^{\mu}$ constructed using $W_\mu$, and $\mathbf{w}_i^\mu$ is one of its $l_n$ basis vectors. It should be noted that practically the $D^U$ matrices can be generated as:
\begin{equation}
D^U(g) = A(g) \otimes R(g),
\end{equation}
where $A(g)$ is the matrix representing the mapping between atoms in the path for a given symmetry operation $g\in G$, and $R(g)$ is the $3\times3$ matrix representing the proper or improper rotation associated with $g$ in the fractional crystal basis of the system. 

Therefore, for a given reciprocal lattice vector in the first Brillouin zone, and commensurate distortion space (see Supplementary Note 1), $U$ can be decomposed into a finite set of symmetry-invariant irreducible subspaces:
\begin{equation}
U = \bigoplus_\mu m_\mu W_\mu,
\end{equation}
where $m_\mu$ are positive integers indicating the number of times $W_\mu$ appears in $U$.
Consequently, all of the basis vectors ($\mathbf{w}_i^\mu$) are orthogonal, and together form a complete basis for $U$,
\begin{equation}
\mathbf{w}_i^\mu[\mathbf{w}_{i'}^{\mu'}]^\text{T} = \delta_{i,i'}\delta_{\mu,\mu'}.
\end{equation}
Additionally, a similar orthonormality relation exists between irrep matrices,
\begin{equation}
\sum_{g \in G} D_{ij}^{\mu}(g)\left[D_{i'j'}^{\mu'}(g)\right]^* = \dfrac{h}{l_n}\delta_{i,i'}\delta_{j,j'}\delta_{\mu,\mu'},
\label{ortho}
\end{equation}
where $h$ is the number of elements $g \in G$, and $l_n$ is the dimension of the irrep.

These symmetry-adapted basis vectors are prime candidates for perturbations that break some, but not all, distortion symmetry because they have the symmetry that transforms as the kernel of the associated irrep: if $D^{\mu}_{ii'}(g) = \delta_{i,i'}$ for $1\leq i\leq l_n$ (for one dimensional irreps, this simply means that $D^{\mu}(R) = [1]$), $g$ leaves $\mathbf{w}_i^\mu$ unchanged and is thus a symmetry operation of $\mathbf{w}_i^\mu$.

As mentioned, it is important to note that the unstable displacive modes that could be generated for a path would transform as a specific irrep of its distortion symmetry group. Due to this, they would be able to be expressed as a linear combination of the described symmetry-adapted basis vectors of the subspace associated with that particular irrep. As such, perturbing a path with these provides a feasible, systematic, and accurate alternative to perturbing with unstable modes that may exist.


\subsubsection{Generating perturbations with projection operators}

One way to generate symmetry-adapted basis vectors is through the method of projection operators \cite{Dresselhaus2008,Bradley2009}. A projection operator $\hat{P}_{kl}^{\mu}$ is an operator that transforms a basis vector $\mathbf{w}_l^\mu$ corresponding to the symmetry invariant subspace associated with $D^\mu$ into another basis vector $\mathbf{w}_k^\mu$ of the same subspace\cite{Dresselhaus2008}:
\begin{equation}
\hat{P}_{kl}^{\mu} \mathbf{w}_l^\mu = \mathbf{w}_k^\mu
\label{proj_app}
\end{equation}
To obtain these operators, let the projection operator be written as the linear combination of symmetry operations $g$:
\begin{equation}
	\hat{P}_{kl}^{\mu} = \dfrac{l_n}{h}\sum_{g \in G} A_{kl}(g) g \label{proj_coeff}
\end{equation}
where $A_{kl}(g)$ is a constant. By substituting Eq.~\ref{proj_coeff} into Eq.~\ref{proj_app}, we obtain:
\begin{equation}
\dfrac{l_n}{h}\sum_{g \in G} A_{kl}(g) g \mathbf{w}_l^\mu = \mathbf{w}_k^\mu.
\label{proj_sub1}
\end{equation}
Multiplying both sides by $\left[\mathbf{w}_k^\mu\right]^\text{T}$ then yields
\begin{eqnarray}
\dfrac{l_n}{h}\sum_{g \in G} A_{kl}(g)\left[\mathbf{w}_k^\mu\right]^\text{T} g \mathbf{w}_l^\mu &=& \left[\mathbf{w}_k^\mu\right]^\text{T}\mathbf{w}_k^\mu
\\
\dfrac{l_n}{h}\sum_{g \in G} A_{kl}(g)D^{\mu}_{kl}(g) &=& 1.
\end{eqnarray}
Using the orthonormality relation in Eq.~\ref{ortho}, $A_{kl}(g)$ can then be written as:
\begin{equation}
A_{kl}(g) = \dfrac{l_n}{h}\left[D^{\mu}_{kl}(g)\right]^*.
\end{equation}
This finally allows us to write the projection operator $\hat{P}_{kl}^{\mu}$ as:
\begin{equation}
\hat{P}_{kl}^{\mu} = \dfrac{l_n}{h}\sum_{g \in G} \left[D^{\mu}_{kl}(g)\right]^* g \label{proj_op}
\end{equation}
It is important to note that applying the projection operator $\hat{P}_{kk}^{\mu}$ (which maps to $\mathbf{w}_k^\mu$ to itself) to an arbitrary vector generates the orthogonal projection of said vector onto $\mathbf{w}_k^\mu$. For irreps with $l_n > 1$, we can also generate the other basis vectors by applying $\hat{P}_{kl}^{(\Gamma_n)}$ to $\mathbf{w}_l^\mu$ once it is projected out of an arbitrary vector. 

To generate a symmetry adapted perturbation to a path with a distortion symmetry group $G$ that transforms as irrep $D^{\mu}$, $\hat{P}_{kk}^{\mu}$ is applied to the basis vectors of distortion space. This will project out of each basis vector the component of the symmetry adapted basis vector $\mathbf{w}_k^{\mu}$ of $W_\mu \subset U$:
\begin{equation}
\hat{P}_{kk}^{\mu} \mathbf{d}_i^j = c^j_i\mathbf{w}_{k}^{\mu},
\end{equation}
where $c^j_i$ is a constant.

By applying the operator to every basis vector of $U$, all basis vectors of $W_{\mu}$ can be generated. However, repeats of symmetry-adapted basis vectors are likely to be generated, from which one of each linearly independent vector can be chosen to construct the perturbation. For subspaces of dimension larger than one, or for multiple subspaces that may exist in a particular $D^U$, a perturbation created from the linear combination of the basis vectors with different random coefficients will bring the path symmetry down to that of the kernel of the irrep. However, in the former case, path symmetries corresponding to epikernels may be imposed with specific weights for each vector, which can also be exploited to explore higher symmetries if desired. 

It should be noted that the sum in Eq.~(\ref{proj_op}) that runs over all group elements $g\in G$ is infinite for distortion groups that include translational symmetry for periodic structures. This is circumvented by choosing a unit cell of the structure, imposing periodic boundary conditions by using a fractional coordinate system, and only summing over symmetry elements within that cell. Depending on the irrep that is chosen to construct the perturbation, the cell choice becomes an important consideration. For irreps at $\mathbf{b}\neq(0,0,0)$, where $\mathbf{b}$ is a reciprocal lattice vector in the first Brillouin zone (see Supplementary Note 1), perturbations will result in a loss of translational symmetry, and a cell must be chosen such that this is accommodated. To help with this, we can use the fact that the elements $D^{\mu}(g)_{kk}$ can be written as
\begin{equation}
D^{\mu}_{kk}(g) = e^{-i\mathbf{b}\boldsymbol{\cdot}\mathbf{A}} D^{\mu}_{kk}(r)
\end{equation}
where $r$ consists of the rotation and nonelementary translational parts of the element $g\in G$, and $\mathbf{A}$ is a vector of the elementary translational component\cite{Kovalev1964,Koster1957,Bradley2009}. Once $\mathbf{b}$ is chosen, if the term $e^{-i\mathbf{b}\boldsymbol{\cdot}\mathbf{A}} \neq 1$ for a particular translation vector $\mathbf{A}$, and for all $\mathbf{b}$ in the star of the group of the wave vector (see Supplementary Note 1), it is those symmetries that will be broken for perturbations constructed with any irrep associated with $\mathbf{b}$. In response, a supercell should be chosen that explicitly includes atoms generated from elements with elementary translations $\mathbf{A}$. 

\section{Implementation and Examples}
To enable the use of the methods described thus far, the projection operator approach to generating the symmetry-adapted perturbations was implemented into a Python package (\texttt{DiSPy})\cite{dispy}. This utility has been made to interface with many of the most popular software packages that support the NEB method, and allows for the generation of images describing perturbed initial paths.

To identify standard spatial symmetries from which the path symmetry group is generated, the well documented and open-source \texttt{SPGLIB} package\cite{A.2009} is utilized. In order to employ projection operators, the matrix representation of all distortion group elements for any given irrep is required. Since non-magnetic distortion groups are isomorphic to space groups, the listing containing irrep matrices for all of the space groups by Stokes \textit{et. al.}\cite{Stokes2013} is used.

\begin{figure}
	\centerline{\includegraphics[width=2.6in]{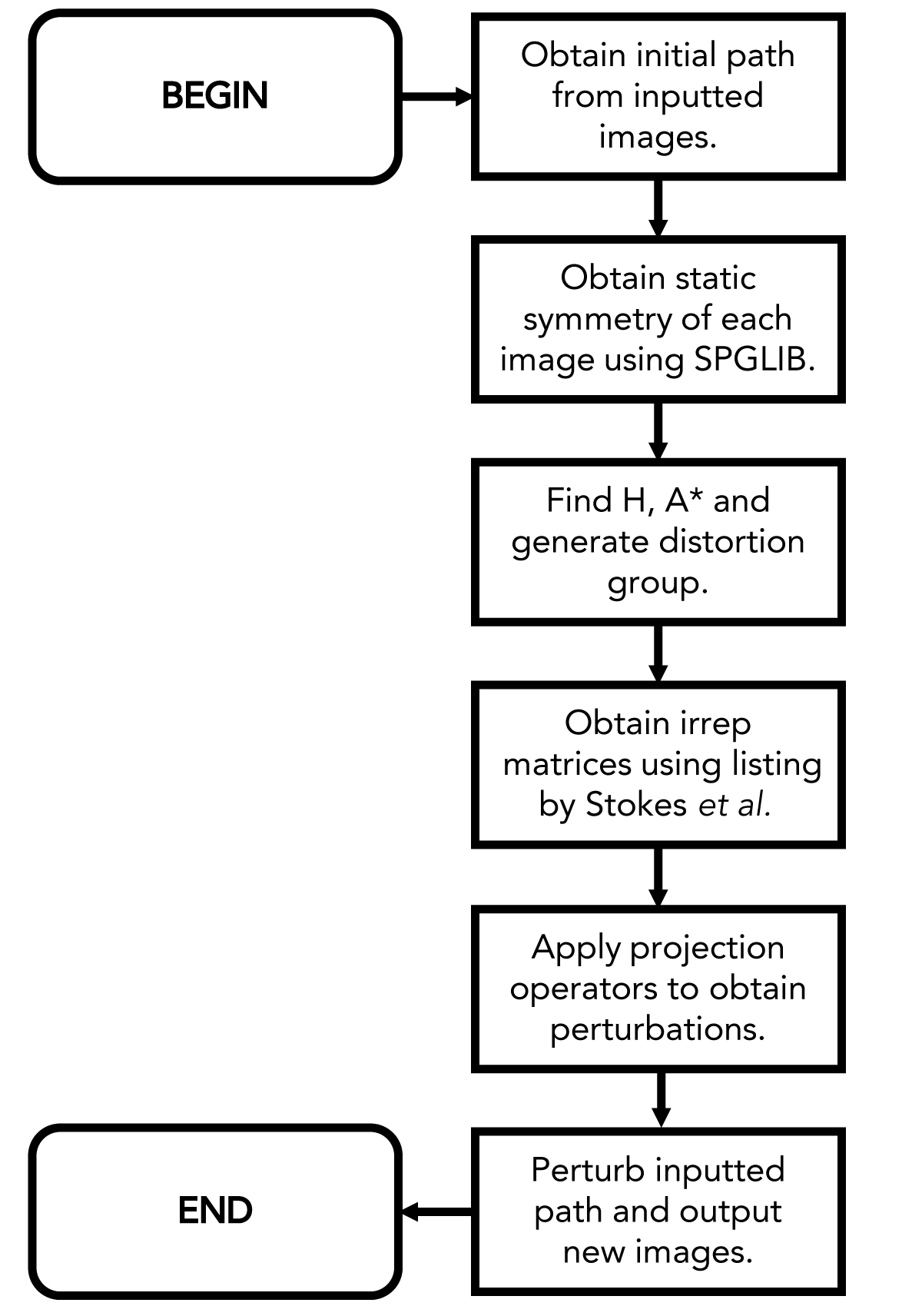}} 
	\caption{Flowchart illustrating the steps taken by \texttt{DiSPy} to generate perturbed initial paths for the NEB algorithm. \label{chart}}
\end{figure}

A flowchart illustrating the function of \texttt{DiSPy} is presented in Figure \ref{chart}. First, the elements of the distortion group of an inputted initial path are generated from the spatial symmetry elements of the space groups of the individual images obtained using \texttt{SPGLIB}. These consist of matrix-vector pairs representing the symmetry operations of the group in the fraction crystal basis of the inputted structures. Next, these matrices and vectors are transformed into the standard basis of the isomorphic space group as defined in the International Tables of Crystallography (ITA)\cite{Hahn2006}. Using the specific irrep designated by the user, the irrep matrices can then be looked up using these standardized matrix-vector pairs. From here, symmetry-adapted perturbations for the given irrep are generated using projection operators and the basis vectors of the distortion space. These are subsequently applied to the images of the path. Finally, The elements of the new distortion symmetry group are obtained, and the perturbed images are outputted.

\subsection{LiNbO$_3$ ferroelectric switching}

To the illustrate the described procedure, NEB calculations are applied to the study of ferroelectric switching pathways in LiNbO$_3$, with the first example consisting of uniform bulk switching. Although this provides a simplistic model of how switching will proceed in the material, it can still produce useful insight into the kinds of atomic motion that may be involved. 

\begin{figure}
	\centerline{\includegraphics[width=3.0in]{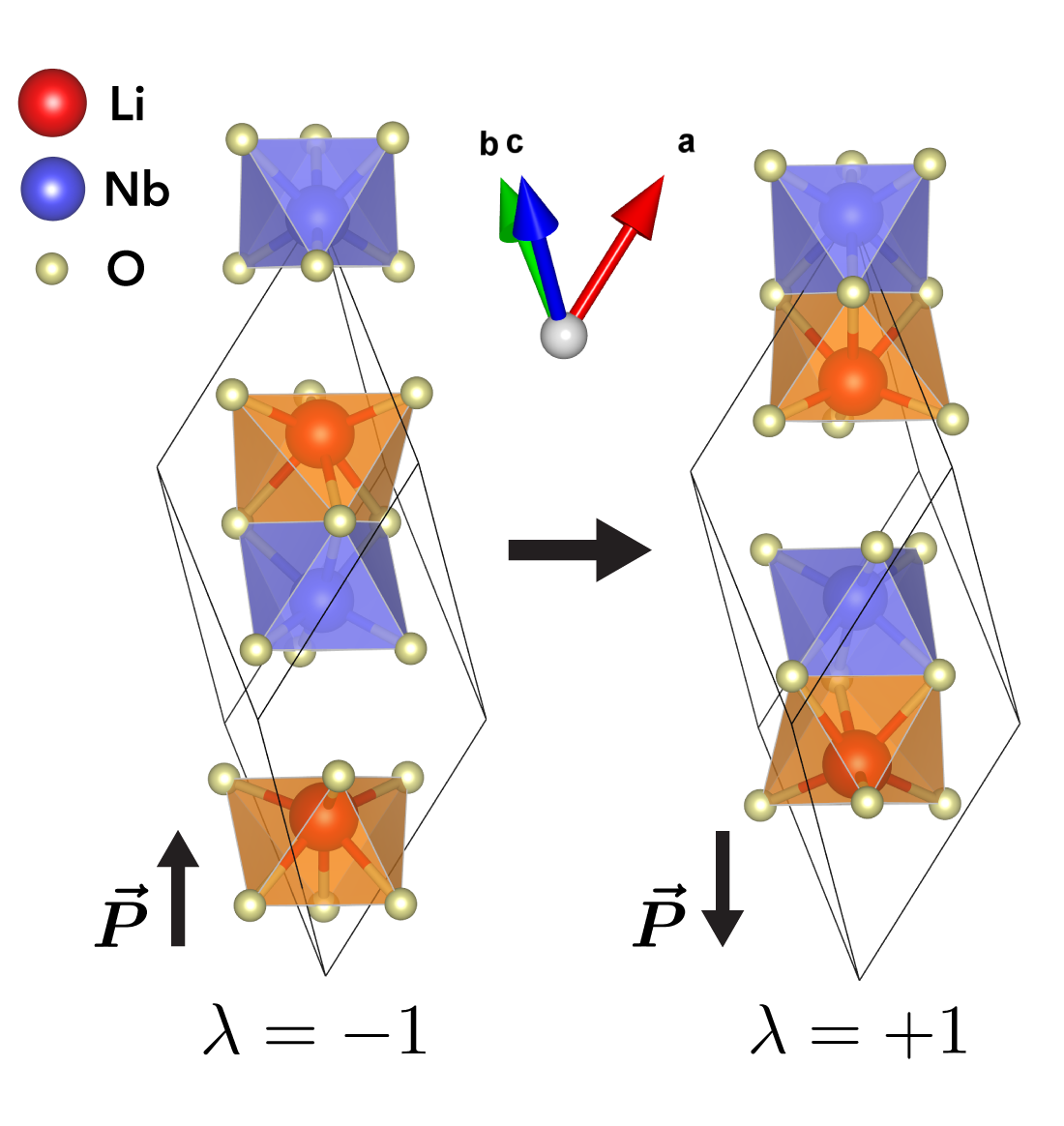}} 
	\caption{The initial and final states of the primitive unit cell for bulk polarization switching in LiNbO$_3$. Arrows indicate the direction of the polarization along $[111]$. The switching process consists of coordinated motion along this three-fold axis of the Li cations as they move from one oxygen octahedra to another.\label{lno_unit}}
\end{figure}

LiNbO$_3$ is a ferroelectric material with wide use in optical, electro-optical, and piezoelectric applications \cite{Inbar1997}.  Below \SI{1480}{\kelvin}, it exists in a polar ground state with rhombohedral symmetry ($R3c$), and above, in a non-polar paraelectric phase ($R\bar{3}c$) \cite{Veithen2002}. The origin of its ferroelectricity can be attributed to a polar mode associated with the Li cations being displaced along the three-fold rhombohedral axis. Reversing the polarization of the structure then involves the motion of these cations along this axis from one oxygen octahedra to another (see Figure \ref{lno_unit}). This switching behavior has previously been explored using density functional theory calculations by Ye and Vanderbilt \cite{Ye2016}. Due to the simplicity of the system, two options for a coherent switching pathway were put forward and studied: one in which the two Li cations move simultaneously (this is obtained by a simple linear interpolation between the images in Figure \ref{lno_unit}), and one in which they move sequentially for at least some portion of the path. For the pathway resulting from simultaneous motion, the intermediate image is that of the high-temperature paraelectric structure ($R\bar{3}c$), and for the sequential pathway, it is a lower symmetry structure with $R\bar{3}$ symmetry \cite{Ye2016}. In order to calculate the energy profile of the sequential path, DFT calculations were run to obtain the total energy of the static structures at various points along the reaction pathway \cite{Ye2016}. To incorporate the sequential motion of the Li cations into the calculations, structural optimizations of the static images were completed with the Li atoms frozen in place at various positions along the three-fold axis. The resulting sequential path showed a lower energy profile than that of the path with simultaneous motion, revealing the surprising fact that the midpoint of the switching path is not identified with the high-symmetry paraelectric structure of $R\bar{3}c$.

\begin{figure*}
	\centerline{\includegraphics[width=\linewidth]{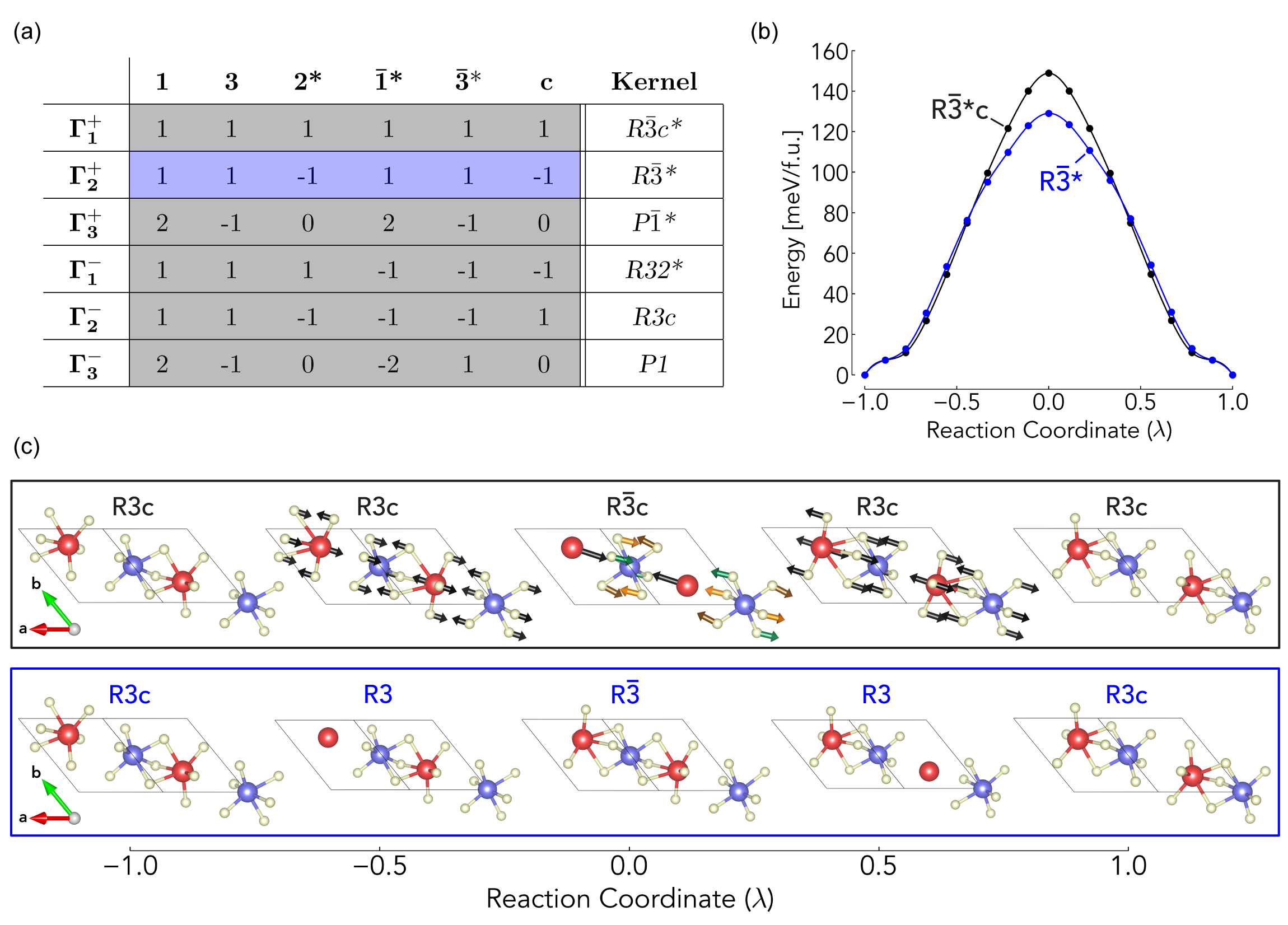}} 
	\caption{Path perturbations and NEB data for bulk ferroelectric switching in LiNbO$_3$. (a) Character table of the distortion symmetry group of the initial simultaneous path ($R\bar{3}^*c$) at $\vec{b}=(0,0,0)$. Perturbations constructed for a specific irrep will reduce the path symmetry to that of the kernel shown in the last collumn. (b) The energy relative to the initial and final state as a function of reaction coordinate for the simultaneous ($R\bar{3}^*c$) and sequential ($R\bar{3}^*$) pathways obtained from NEB calculations. The sequential path is obtained by perturbing the initial relaxed path using symemtry-adapted perturbations constructed from the $\Gamma_2^{+}$ irrep. (c) Snapshots of images along the relaxed simultaneous and sequential pathways given by the NEB algorithm. The $R\bar{3}^*c$ and $R\bar{3}^*$ pathways are illustrated by a black and blue outline respectively. The space group of each image is shown above it. Perturbations to the initial $R\bar{3}^*c$ path are shown for the atoms involved in the distortion (Li and O). Black, brown, green, and orange arrows indicate perturbations along $[111]$, $[121]$, $[112]$, and $[211]$ respectively.\label{lno_dat1}}
\end{figure*}

Although the above result was obtained without the use of the NEB method, the problem can be simplified by its use alongside the consideration of distortion symmetry. Choosing the simultaneous (linear interpolated) path as the starting point for an NEB calculation, the high energy profile shown in Fig.~\ref{lno_dat1}b is produced. As mentioned, since the NEB algorithm cannot break the distortion symmetry of a path, the group of the path is conserved throughout the calculation. This, in turn, makes the sequential path of lower energy inaccessible without perturbation. To find the distortion group, the space group of each of the images is obtained. These are listed above each of the image illustrations for the linear path outlined in black in Fig.~\ref{lno_dat1}c. Next, the elements of $H$ and $A$ can be found as,
\begin{eqnarray}
&H&  = \{1,3^{+}_{111},3^{-}_{111},n_{\bar{1}01},n_{1\bar{1}0},n_{01\bar{1}}\}\\
&A&  = \{\bar{1},2_{\bar{1}01},2_{1\bar{1}0},2_{01\bar{1}},\bar{3}_{111}^{+},\bar{3}_{111}^{-}\},
\end{eqnarray}
written in ITA\cite{Hahn2006} notation in the rhobohedral cell basis. Finally, the distortion group can be obtained:
\begin{equation}
G = H \cup 1^* A = R\bar{3}^*c
\end{equation}
By considering the non-trivial irreps of the group $R\bar{3}^*c$, symmetry-adapted perturbations to the initial relaxed path can be constructed and applied. Since the path images contain just the primitive rhombohedral cell, only non-trivial irreps at $\Gamma$, $\mathbf{b} = (0,0,0)$, can be used, as no perturbations that break translational symmetry can be accommodated. The characters of the matrices in these irreps can be seen in the character table shown in Fig.~\ref{lno_dat1}a. The full irrep matrices can be found using the listing by Stokes \textit{et al.}, and projection operators can then be used to generate perturbations to the path. For most of the perturbations (those constructed using the irreps highlighted in black in Fig.~\ref{lno_dat1}a), running NEB calculation simply returns the path back to the path with a distortion group of $R\bar{3}^*c$. However, for the new initial path resulting from perturbing with symmetry-adapted basis vectors constructed with the $\Gamma_{2+}$ irrep (see Fig.~\ref{lno_dat1}c), the lower energy sequential path is obtained. This path has a distortion symmetry group of $R\bar{3}^*$, which is equal to that of the kernel of $\Gamma_{2+}$ (Fig.~\ref{lno_dat1}a). From here, further perturbations are able to be applied to the newly obtained path. Using the irreps of the groups $R\bar{3}^*$, symmetry-adapted perturbations are constructed and applied to the respective relaxed paths. NEB calculations are then run resulting in all final paths returning to the initial $R\bar{3}^*$. This indicates that there are no additional lower energy paths of similar character to be found. 
%

\begin{figure*}
	\centerline{\includegraphics[width=\linewidth]{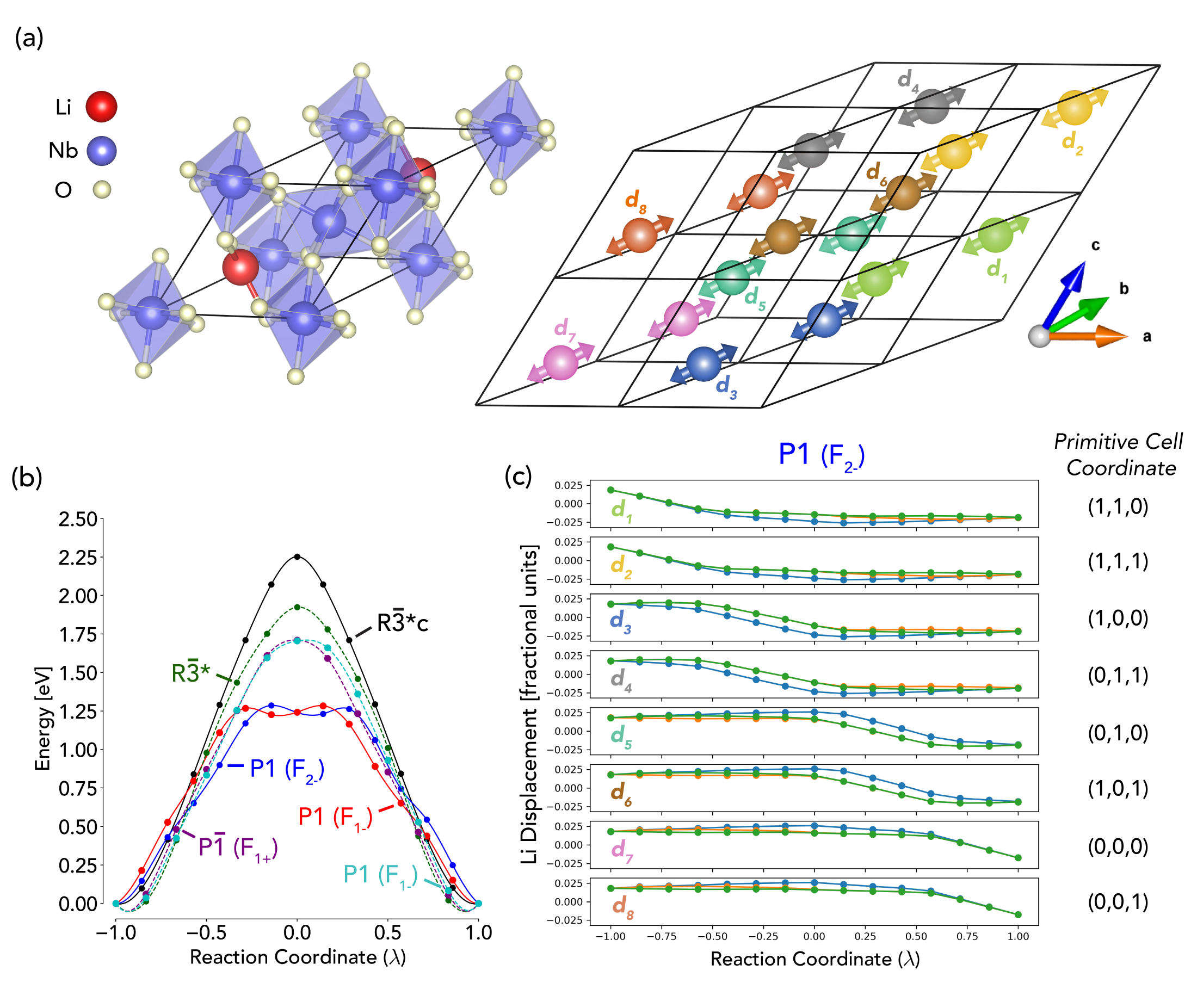}} 
	\caption{Path perturbations using other reciprocal lattice vectors. (a) Illustration of the unit cell and $2\times2\times2$ supercell of LiNbO$_3$. The high symmetry structure with $R\bar{3}c$ symmetry is shown, along with the positions of all the Li atoms in each of the eight primitive rhombohedral unit cells. Arrows and labels indicate atomic displacements of Li from the high-symmetry positions. (b) The energy relative to the initial and final state as a function of reaction coordinate for the simultaneous ($R\bar{3}^*c$) and sequential ($R\bar{3}^*$) paths, as well as the lower energy paths from their perturbation. Paths found by perturbing the simultaneous and sequential paths are shown with solid and dotted lines respectively. The perturbations were obtained using symmetry-adapted basis vectors constructed from the $F_{2-}$ and $F_{1-}$ irreps of $R\bar{3}^*c$ and the $F_{1-}$ and  $F_{1+}$ irreps of $R\bar{3}^*$. (c) Plot of the displacement of both Li atoms within each primitive rhombohedral unit cell along each of the cell vectors in panel (a) as a function of reaction coordinate for the blue path in panel (b). The high-symmetry structure with $R\bar{3}c$ symmetry is taken as reference. What is illustrated is a switching process with a lower overall energy barrier that involves the step wise transition of different pairs of primitive cells.\label{lno_dat2}}
\end{figure*}

In order to construct perturbations using other high-symmetry $\mathbf{b}$-vectors, a $2\times2\times2$ rhombohedral supercell can instead be used for all of the images in the initial $R\bar{3}^*c$ path. In turn, this allows for the exploration of the potential ways in which coordinated motion between different unit cells can reduce cost of switching. 

An illustration of the supercell can be seen in Fig.~\ref{lno_dat2}a. Running NEB calculations on the initial $R\bar{3}^*c$ path results in the energy profile shown in black in Fig.~\ref{lno_dat2}b. Choosing the $F$ reciprocal lattice point at $\mathbf{b} = (1/2,1/2,0)$, path perturbations can then be generated. The loss of translational symmetry can be identified by the three $\mathbf{b}$-vectors in the star of the wave-vector ${\mathbf{b}=(1/2,1/2,0),(0,1/2,1/2),(1/2,0,1/2)}$ (see Supplementary Note 1). Since only unit translation vectors of $\mathbf{A} = (0,0,0)$ and $\mathbf{A} = (1,1,1)$ result in $e^{-i\mathbf{b}\boldsymbol{\cdot}\mathbf{A}} = 1$, all other unit translations in the supercell will be broken after perturbation. Furthermore, it is important to note that all irreps associated with the $F$-point are three-dimensional. Consequently, three symmetry-adapted basis vectors will be generating from applying the projection operators, which in linear combination will make up the final path perturbation. Since we would like to perturb to the kernel symmetry of each irrep, a random and different coefficient for each is chosen.

After constructing perturbations with both $F$-point irreps, two new lower energy paths are obtained that both have trivial path symmetry groups ($P1$). Both are similar in character, and their energy profiles can be seen in Fig.~\ref{lno_dat2}b. In each case, the switching process proceeds in a stepwise manor, with different pairs of unit cells switching simultaneously. In both paths, the two Li atoms in each primitive cell move together, and their displacement from their positions in the high-symmetry $R\bar{3}c$ structure is plotted in Fig.~\ref{lno_dat2}c for the path resulting from perturbing with the $F_{2-}$ irrep. The same procedure has also been completed for the sequential path with $R\bar{3}^*$ symmetry. Two paths with lower energy than the initial path are also found, and their energy profiles are plotted in Fig. \ref{lno_dat2}b. Similar coordinated motion between Li atoms across primitive cells can be seen, but instead with four of the eight cells transitioning at a time. The Li displacements have also been plotted and can be seen in Supplementary Figs. 2-3. Overall, the exhibited coordinated motion between different primitive cells shown in all of the obtained paths allows for a lowering of the maximum switching barrier, and may inform how nucleation and growth switching processes could begin at the unit cell scale at domain walls in the material.

\section{Conclusion}
In this paper, we have presented the mathematical formalism of the distortion symmetry method described by Munro \textit{et al.}\cite{Munro2018}. By considering the distortion space for a given path, projection operators can be applied to generate symmetry-adapted basis vectors of its symmetry-invariant subspaces. These basis vectors can the be used to perturb a path, reducing its symmetry to that of the kernel of the irreducible representation associated with a particular subspace. In turn, this allows one to break free of the distortion symmetry conservation exhibited by the NEB algorithm, induce path instabilities, and explore additional low energy paths that may exist. The described procedure has been implemented into a Python package (\texttt{DiSPy})\cite{dispy}, and has been applied to bulk ferroelectric switching in LiNbO$_3$. Previously reported paths are recreated, with additional low-energy paths found that involve the breaking of translation symmetry. These provide insight into the coordinated motion of atoms across unit cells that may be involved in the switching process. We foresee the generation and application of symmetry adapted perturbations being an integral part of NEB calculations in the future. Furthermore, we envision possible extensions to the distortion symmetry framework that involve other types of symmetry such as distortion translation\cite{Padmanabhan2017,Liu2018}. 

\section*{Methods}
All NEB \cite{JONSSON1998} calculations were completed using the Vienna Ab Initio Simulation Package (VASP) \cite{Kresse1993,Kresse1996,Kresse1996a,Kresse1999} after obtaining the optimized geometries of the end state structures. All perturbations were generated and applied using the \texttt{DiSPy} package \cite{dispy}, and were normalized such that the maximum displacement of any one atom along any of the rhombohedral cell vector directions was set to 0.05\AA. 

All first-principles calculations were completed using the revised Perdew-Burke-Ernzerhof generalized-gradient approximation functional \cite{Perdew2008} (PBEsol) that has been shown to improve the properties of densely packed solids. A 6$\times$6$\times$6 and 3$\times$3$\times$3 $k$-point mesh was used for the unit cell and supercell calculations respectively. A \si{600}{eV} plane-wave cutoff was used for all calculations, with an energy error threshold of \SI{1e-6}{\electronvolt} and \SI{1e-5}{\electronvolt} being used for the geometry optimization and NEB calculations respectively. Both types of calculations were run until forces were below \SI{0.001}{\electronvolt\per\angstrom} and \SI{0.01}{\electronvolt\per\angstrom} respectively. The projector augmented wave method was used to represent the ionic cores. There were 3 electrons for Li ($1s^22s^2$), 13 electrons for Nb ($4s^24p^64d^45s^1$), and 6 electrons for O ($2s^22p^4$) treated explicitly.

\section*{Data Availability}
The data that supports the findings of this study are available from the corresponding author upon request.

\section*{Acknowledgments}
This material is based upon work supported by the National Science Foundation under Grants No. 1807768 and No. 1210588. We acknowledge the support of the Natural Sciences and Engineering Research Council of Canada (NSERC), and the NSF-MRSEC Center for Nanoscale Science at the Pennsylvania State University, Grant No. DMR-1420620. J.M.M. and I.D. also acknowledge partial support from the Soltis faculty support award and the Ralph E. Powe junior faculty award from Oak Ridge Associated Universities.

\vspace{-0.3cm}

\section*{Author Contributions}
The implementation of the distortion symmetry method was completed by J.M. and V.L. All calculations presented were completed by J.M. The manuscript was written by J.M., V.S., and I.D. All authors discussed the results and implications, and commented on the manuscript at all stages.
\\

\noindent \textbf{Competing interests:} The authors declare no competing financial or non-financial interests.

\vfill

\bibliography{ref}

\end{document}